\documentclass[12pt,a4paper]{{amsart}}
\usepackage{verbatim} 
\usepackage{amssymb,amsfonts,latexsym,amscd}
\newcommand{\di}{\,\mathrm{d}}
\newcommand{\R}{\mathbb{R}}
\newcommand{\V}{\tilde V}
\newcommand{\Hn}{\mathbb H^n}

\newtheorem{Theorem}{Theorem}[section]
\newtheorem{Lemma}{Lemma}[section]
\newtheorem{Fact}{Fact}[section]
\newtheorem{Corollary}{Corollary}[section]

\author{Rafael D. Benguria \and Helmut Linde} %
\email{RBenguri@fis.puc.cl, Helmut.Linde@gmx.de}
\title[A second eigenvalue bound]
{A second eigenvalue bound for the Dirichlet Schr\"odinger operator}
\address{Department of Physics, Pontific\'\i a Universidad Cat\'olica de Chile Casilla 306, Correo 22
Santiago, Chile.}
\thanks{R.B. was supported by FONDECYT project \# 102-0844. H.L. gratefully acknowledges financial support from DIPUC of the Pontif\'\i cia Universidad Cat\'olica de Chile and from CONICYT}

\begin{document}

\begin{abstract}
Let $\lambda_i(\Omega,V)$ be the $i$th eigenvalue of the Schr\"odinger operator with Dirichlet boundary conditions on a
bounded domain $\Omega \subset \R^n$ and with the positive potential $V$. Following the spirit of the
Payne-P\'olya-Weinberger conjecture and under some convexity assumptions on the spherically rearranged potential
$V_\star$, we prove that $\lambda_2(\Omega,V) \le \lambda_2(S_1,V_\star)$. Here $S_1$ denotes the ball, centered at the
origin, that satisfies the condition $\lambda_1(\Omega,V) = \lambda_1(S_1,V_\star)$.

Further we prove under the same convexity assumptions on a spherically symmetric potential $V$, that $\lambda_2(B_R, V)
/ \lambda_1(B_R, V)$ decreases when the radius $R$ of the ball $B_R$ increases.

We conclude with several results about the first two eigenvalues of the Laplace operator with respect to a measure of
Gaussian or inverted Gaussian density.
\end{abstract}
\maketitle

\section{Introduction}
In an earlier publication \cite{AB2}, Ashbaugh and one of us have proven the Payne-P\'olya-Weinberger (PPW) conjecture,
which states that the first two  eigenvalues $\lambda_1, \lambda_2$ of the Dirichlet-Laplacian on a bounded domain
$\Omega\subset \mathbb R^n$ ($n\ge 2$) obey the bound
\begin{equation} \label{EqPPW}
\lambda_2 / \lambda_1 \le j_{n/2,1}^2/j_{n/2-1,1}^2.
\end{equation}
Here $j_{\nu,k}$ stands for the $k$th positive zero of the Bessel function $J_\nu$. Thus the right hand side of
(\ref{EqPPW}) is just the ratio of the first two eigenvalues of the Dirichlet-Laplacian on an $n$-dimensional ball of
arbitrary radius. This result is optimal in the sense that equality holds in (\ref{EqPPW}) if and only if $\Omega$ is a
ball.

The proof of the PPW conjecture has been generalized in several ways. In \cite{AB1} a corresponding theorem has been
established for the Laplacian operator on a domain $\Omega$ that is contained in a hemisphere of the $n$-dimensional
sphere $\mathbb S^n$. More precisely, it has been shown that $\lambda_2(\Omega) \le \lambda_2(S_1)$, where $S_1$ is the
$n$-dimensional geodesic ball in $\mathbb S^n$ that has $\lambda_1(\Omega)$ as its first Dirichlet eigenvalue.

A further variant of the PPW conjecture has been considered by Haile. In \cite{H} he compares the second eigenvalue
$\lambda_2(\Omega,kr^\alpha)$ of the Schr\"o\-din\-ger operator with the potential $V = kr^\alpha$ ($k>0, \alpha\ge 2$)
with $\lambda_2(S_1,kr^\alpha)$, where $S_1$ is the ball, centered at the origin, that satisfies the condition
$\lambda_1(\Omega,kr^\alpha) = \lambda_1(S_1,kr^\alpha)$. Here and in the following we denote by $\lambda_i(\Omega,V)$
the $i$th eigenvalue of the Schr\"o\-din\-ger operator $-\Delta + V(\vec r)$ with Dirichlet boundary conditions on a
bounded domain $\Omega \subset \R^n$.

We have to mention a gap in \cite{H}, which occurs in the proof of Lemma 3.2. The author claims (and uses) that all
derivatives of the function $Z(\theta)$ (which is equal to $T'(\theta)$ where $T(\theta)=0$) coincide with the
derivatives of $T'(\theta)$ in the points where $T(\theta)=0$. This is not proven and there seems to be no reason why
it should be true. The same problem occurs in the proof of Lemma 3.3. In the present paper we will prove a theorem that
includes Haile's theorem as a special case and thus remedies the situation.

One very important difference between the original PPW conjecture and the extended problems in \cite{AB1,H} is that in
the later cases the ratio $\lambda_2/\lambda_1$ is not scaling invariant anymore. While $\lambda_2/\lambda_1$ is the
same for any ball in $\mathbb R^n$, it is an increasing function of the radius for balls in $\mathbb S^n$ \cite{AB1}.
On the other hand, we will see that $\lambda_2(B_R,V)/\lambda_1(B_R,V)$ on the ball $B_R$ is a decreasing function of
the radius $R$, if $V$ has certain convexity properties. This rises the question which is the `right size' of the
comparison ball in the PPW estimate. We will make some remarks on this problem below.

The main objective of the present work is to prove a PPW type result for a Schr\"odinger operator with a positive
potential. We will state the corresponding theorem in the following section. In Section \ref{SectionGauss} we will
transfer our results to the case of a Laplacian operator with respect to a metric of Gaussian or inverted Gaussian
measure, the two cases of which are closely related to the harmonic oscillator. The rest of the article will be devoted
to the proofs of our results.

\section{Main Results} \label{SectionResults}

Let $\Omega \subset \R^n$ (with $n\ge 2$) be some bounded domain and $V : \Omega \rightarrow \R^+$ some positive
potential such that the Schr\"odinger operator $-\Delta + V$ (subject to Dirichlet boundary conditions) is self-adjoint
in $L^2(\Omega)$. We call $\lambda_i(\Omega,V)$ its $i$th eigenvalue. Further, we denote by $V_\star$ the radially
increasing rearrangement of $V$. Then the following PPW type estimate holds:
\begin{Theorem} \label{TheoremPPW}
Let $S_1 \subset \R^n$ be a ball centered at the origin and of radius $R_1$ and let $\V: S_1 \rightarrow \R^+$ be some
radially symmetric positive potential  such that $\V(r) \le V_\star(r)$ for all $0 \le r \le R_1$ and
$\lambda_1(\Omega,V) = \lambda_1(S_1,\V)$. If $\V(r)$ satisfies the conditions
\begin{enumerate}
\item[a)] $\V(0) = \V'(0)=0$ and
\item[b)] $\V'(r)$ exists and is increasing and convex,
\end{enumerate}
then
\begin{equation}\label{EqTheorem}
\lambda_2(\Omega,V) \le \lambda_2(S_1,\V).
\end{equation}
\end{Theorem}
If $V$ is such that $V_\star$ satisfies the convexity conditions stated in the theorem, the best bound is obtained by
choosing $\V = V_\star$. In this case the theorem is a typical PPW result and optimal in the sense that equality holds
in (\ref{EqTheorem}) if $\Omega$ is a ball and $V = V_\star$. For a general potential $V$ we still get a non-trivial
bound on $\lambda_2(\Omega,V)$ though it is not sharp anymore. To show that our Theorem \ref{TheoremPPW} contains
Haile's result \cite{H} as a special case, we state the following corollary:
\begin{Corollary}
Let $\V: \R^n \rightarrow \R^+$ be a radially symmetric positive potential that satisfies the conditions a) and b) of
Theorem \ref{TheoremPPW} and let $S_1 \subset \R^n$ be the ball (centered at the origin) such that
$\lambda_1(\Omega,\V) = \lambda_1(S_1,\V)$. Then
$$\lambda_2(\Omega,\V) \le \lambda_2(S_1,\V).$$
\end{Corollary}

The proof of Theorem \ref{TheoremPPW} follows the lines of the proof in \cite{AB2} and will be presented in Section
\ref{SectionProof}. Let us make a few remarks on the conditions that $\V$ has to satisfy. Condition a) is not a very
serious restriction, because any bounded potential can be shifted such that $V_\star(0) = 0$. Also $V_\star'(0) = 0$
holds if $V$ is somewhat regular where it takes the value zero. Moreover, our method relies heavily on the fact that
\begin{equation} \label{EqLambda}
\lambda_2(B_R,\V) \ge \left(1+\frac 2n\right) \lambda_1(B_R,\V),
\end{equation}
which is a byproduct of our proof and holds for any ball $B_R$ and any potential $\V$ that satisfies the conditions of
Theorem \ref{TheoremPPW}. The conditions a) and b) will be needed to show the above inequality, which is equivalent to
$q''(0) \le 0$ for a function $q$ to be defined in the proof. Numerical studies indicate that b) is somewhat sharp in
the sense that, for example, a potential $r^{2-\epsilon}$ (which violates b) only `slightly') does not satisfy
(\ref{EqLambda}) for every $R$. In this case the statement of Theorem \ref{TheoremPPW} may still be true, but the
typical scheme of the PPW proof will fail. Furthermore, condition a) and b) will allow us to employ the crucial
Baumgartner-Grosse-Martin (BGM) inequality \cite{BGM,AB3}: From a) and b) we see that $V(r) + rV'(r)$ is increasing.
Consequently $rV(r)$ is convex, which is just the condition needed to apply the BGM inequality.

As mentioned above, one has to chose carefully the size of the comparison ball in a PPW estimate if
$\lambda_2/\lambda_1$ is a non-constant function of the ball's radius. In the case of the Laplacian on $\mathbb S^n$,
one compares the second eigenvalues on $\Omega$ and $S_1$, the ball that has the same first eigenvalue as $\Omega$. By
the Rayleigh-Faber-Krahn (RFK) inequality for $\mathbb S^n$ it is clear that $S_1 \subset \Omega^\star$, where
$\Omega^\star$ is the spherically symmetric rearrangement of $\Omega$. It has also be shown in \cite{AB1} that
$\lambda_2/\lambda_1$ on a geodesic ball in $\mathbb S^n$ is an increasing function of the ball's radius. One can
conclude from these two facts that in $\mathbb S^n$ an estimate of the type (\ref{EqTheorem}) is stronger than the
inequality
\begin{equation} \label{EqLL}
\lambda_2(\Omega)/\lambda_1(\Omega) \le \lambda_2(\Omega^\star) / \lambda_1(\Omega^\star).
\end{equation}
It has also been argued in \cite{AB3} why the situation is different in the hyperbolic space $\Hn$. Here an estimate of
the type (\ref{EqLL}) is not possible, for the following reason: One can show that $\lambda_2/\lambda_1$ on geodesic
balls in $\Hn$ is a decreasing function of the radius. Now suppose, for example, that $\Omega$ is the ball $B_R$ with
very long and thin tentacles attached to it. Then the first and the second eigenvalue of the Laplacian on $\Omega$ and
$B_R$ are almost the same, while the ratio $\lambda_2/\lambda_1$ on $\Omega^\star$ can be considerably less than on
$B_R$ (and thus on $\Omega$). We will prove a PPW inequality of the type $\lambda_2(\Omega) \le \lambda_2(S_1)$ for
$\Hn$ and the monotonicity of $\lambda_2/\lambda_1$ on geodesic balls in a future publication.

To shed light on the question which is the right type of PPW inequality for the Schr\"odinger operator on $\Omega$, we
state
\begin{Theorem} \label{TheoremMonotonicity}
Let $V: \R^n \rightarrow \R^+$ be a spherically symmetric potential that satisfies the conditions of Theorem
\ref{TheoremPPW}, i.e.
\begin{enumerate}
\item[a)] $V(0) = V'(0)=0$ and
\item[b)] $V'(r)$ exists and is increasing and convex.
\end{enumerate}
Then the ratio
$$\frac{\lambda_2(B_R, V)}{\lambda_1(B_R, V)}$$
is a decreasing function of $R$.
\end{Theorem}
This theorem shows that one can not replace equation (\ref{EqTheorem}) in our Theorem \ref{TheoremPPW} by an inequality
of the type (\ref{EqLL}), following the same reasoning as in the case of the Laplacian on $\mathbb H^n$. Theorem
\ref{TheoremMonotonicity} will be proven in Section \ref{SectionProof2}.

\section{Connection to the Laplacian operator in Gaussian space} \label{SectionGauss}

Recently, there has been some interest in isoperimetric inequalities in $\R^n$ endowed with a measure of Gaussian ($\di
\mu_- =  e^{-r^2/2}\di^nr$) or inverted Gaussian ($\di \mu_+ =  e^{+r^2/2}\di^nr$) density. For the Gaussian space it
has been known for several years that a classical isoperimetric inequality holds. Yet the ratio of Gaussian perimeter
and Gaussian measure is minimized by half-spaces instead of spherical domains \cite{B}. The `inverted Gaussian' case,
i.e., $\R^n$ with the measure $\mu_+$, is more similar to the Euclidean case: It has been shown recently that a
classical isoperimetric inequality holds and that the minimizers are balls centered at the origin \cite{MPB}.

We consider the Dirichlet-Laplacians $-\Delta_\pm$ on $L^2(\Omega,\di\mu_\pm)$, where $\Omega\varsubsetneqq\R^n$ is a
domain of finite measure $\di\mu_\pm(\Omega)$. These two operators are defined by their quadratic forms
\begin{equation} \label{EqQuadForm}
h_\pm[\Psi] = \int_\Omega |\nabla \Psi(\vec r)|^2 \di \mu_\pm, \quad \Psi \in W^{1,2}_0(\Omega,\di \mu_\pm).
\end{equation}
The eigenfunctions $\Psi^\pm_i$ and eigenvalues $\lambda^\pm_i(\Omega)$ in question are determined by the the
differential equation
\begin{equation} \label{EqEVProblem}
-\sum\limits_{k=1}^n \frac{\partial}{\partial r_k} \left(e^{\pm r^2} \frac{\partial \Psi^\pm_i}{\partial r_k}\right) =
\lambda^\pm_i(\Omega) e^{\pm r^2} \Psi^\pm_i(\vec r).
\end{equation}
There is a tight connection between the operators $-\Delta_\pm$ on a domain $\Omega$ and the harmonic oscillator
$-\Delta+r^2$ restricted to $\Omega$. Their eigenfunctions and eigenvalues are related by \cite{BCF}
\begin{eqnarray}
\Psi^\pm_i(\vec r) &=& \Psi_i(\vec r) \cdot e^{\mp r^2/2} \quad \textmd{and} \nonumber\\
\lambda^\pm_i(\Omega) &=& \lambda_i(\Omega, r^2) \pm n, \label{EqCon}
\end{eqnarray}
denoting by $\Psi_i$ the Dirichlet eigenfunctions of $-\Delta + r^2$ on $\Omega$.

There is an equivalent of the RFK inequality in Gaussian space \cite{BCF} stating that $\lambda^-_1(\Omega)$ is
minimized for given $\mu_-(\Omega)$ if $\Omega$ is a half-space. The corresponding fact for the `inverted' Gaussian
space is that $\lambda^+_1(\Omega)$ is minimized for given $\mu_+(\Omega)$ by the ball centered at the origin. It can
be seen by the RFK inequality for Schr\"odinger operators \cite{L} in combination with (\ref{EqCon}).

Concerning the second eigenvalue, we will now show what our results from Section \ref{SectionResults} imply for the
operators $-\Delta_\pm$. We state
\begin{Theorem} \label{TheoremMonotonicity2}
For the operator $-\Delta_+$ on a ball $B_R$ of radius $R$ (centered at the origin) the ratio
$\lambda_2^+(B_R)/\lambda_1^+(B_R)$ is a strictly decreasing function of $R$.
\end{Theorem}
In Section \ref{SectionProof3} we will derive Theorem \ref{TheoremMonotonicity2} from Theorem \ref{TheoremMonotonicity}
in a purely algebraic way using only the relation (\ref{EqCon}). Repeating the argument for $\mathbb H^n$ from the
previous section, we see that by Theorem \ref{TheoremMonotonicity2} the best PPW result we can expect to get is
\begin{Theorem} \label{TheoremPPW2}
Be $S_1$ the ball (centered at the origin) that satisfies the condition $\lambda^+_1(S_1) = \lambda^+_1(\Omega)$. Then
\begin{equation*}
\lambda_2^+(\Omega) \le \lambda_2^+(S_1).
\end{equation*}
\end{Theorem}
Theorem \ref{TheoremPPW2} follows immediately from Theorem \ref{TheoremPPW} and (\ref{EqCon}). In the same way we
easily get the corresponding version for $-\Delta_-$:
\begin{Theorem} \label{TheoremPPW3}
Be $S_1$ the ball (centered at the origin) that satisfies the condition $\lambda^-_1(S_1) = \lambda^-_1(\Omega)$. Then
\begin{equation*}
\lambda_2^-(\Omega) \le \lambda_2^-(S_1).
\end{equation*}
\end{Theorem}
Yet in this case it is not clear anymore whether $S_1$ is the optimal comparison ball: First, in contrast to the
`inverted' Gaussian case the ratio $\lambda_2^-(B_R)/\lambda_1^-(B_R)$ is not a decreasing function of $R$ anymore.
This can be seen by comparing the values of  $\lambda_2^-(B_R)/\lambda_1^-(B_R)$ for $R\rightarrow 0$ and $R\rightarrow
\infty$: For small $R$ the ratio is close to the Euclidean value ($\approx 2.539$) while for large $R$ it approaches
infinity (by (\ref{EqCon})). Second, the RFK inequality in Gaussian space states that $\lambda_1^-(\Omega)$ is
minimized by half-spaces, not circles. This means that for general $\Omega$ we don't know whether $\Omega^\star$ is
bigger or smaller than $S_1$. For these differences it remains unclear what is the most natural way to generalize the
PPW conjecture to Gaussian space.

\section{A monotonicity lemma} \label{SectionLemma}

In our proof of Theorem \ref{TheoremPPW} we will need
\begin{Lemma}[Monotonicity of $g$ and $B$] \label{LemmaMonotonicityOfgAndB}
Let $\V$, $S_1$ and $R_1$ be as in Theorem \ref{TheoremPPW} and call $z_1(r)$ and $z_2(r)$ the radial parts (both
chosen positive) of the first two Dirichlet eigenfunctions of $-\Delta + \V$ on $S_1$. Set
\begin{eqnarray*}
g(r) &=& \frac{z_2(r)}{z_1(r)} \quad \textmd{and}\\
B(r) &=& g'(r)^2 + (n-1)\frac{g(r)^2}{r^2}
\end{eqnarray*}
for $0 < r < R_1$. Then $g(r)$ is increasing on $(0,R_1)$ and $B(r)$ is decreasing on $(0,R_1)$.
\end{Lemma}

\begin{proof} \cite{H,AB0}
In this section we abbreviate $\lambda_i = \lambda_i(S_1,\V)$. The functions $z_1$ and $z_2$ are solutions of the
differential equations
\begin{eqnarray}
-z''_{1} - \frac{n-1}{r} z'_{1} + \left(\V -\lambda_1\right) z_{1} &=& 0, \label{EquN1}\\
-z''_{2} - \frac{n-1}{r} z'_{2} + \left(\frac{n-1}{r^2} + \V - \lambda_2\right) z_{2} &=& 0 \nonumber
\end{eqnarray}
with the boundary conditions
\begin{equation}\label{EqBoundary}
z'_{1}(0) = 0, \quad z_{1}(R_1) = 0, \quad z_{2}(0) = 0, \quad z_{2}(R_1) = 0.
\end{equation}
This is assured by the BGM inequality \cite{AB0,BGM}, which is applicable because $r\V$ is convex. As in \cite{AB0} we
define the function
$$q(r) := \frac{rg'(r)}{g(r)}.$$
Proving the lemma is thus reduced to showing that $0 < q(r) < 1$ and
$q'(r) < 0$ for $r \in [0,R]$. Using the definition of $g$ and the equations (\ref{EquN1}), one can show that $q(r)$ is
a solution of the Riccati differential equation
\begin{equation} \label{EqRic}
q' = (\lambda_1-\lambda_2) r + \frac{(1-q)(q+n-1)}r - 2q\frac{z_1'}{z_1}.
\end{equation}
It is straightforward to establish the boundary behavior
$$q(0) = 1, \quad q'(0) = 0, \quad q''(0) = \frac 2n \left(\left(1+\frac 2n\right)\lambda_1 - \lambda_2\right)$$
and
$$q(R_1) = 0.$$
\begin{Fact}  \label{Fact1}
For $0 \le r \le R$ we have $q(r) \ge 0$.
\end{Fact}
\begin{proof}
Assume the contrary. Then there exist two points $0 < r_1 < r_2 \le R_1$ such that $q(r_1) = q(r_2) = 0$ but $q'(r_1)
\le 0$ and $q'(r_2) \ge 0$. If $r_2 < R_1$ then the Riccati equation (\ref{EqRic}) yields
$$0 \ge q'(r_1) = (\lambda_1-\lambda_2) r_1 + \frac{n-1}{r_1} > (\lambda_1-\lambda_2) r_2 + \frac{n-1}{r_2} = q'(r_2)
\ge 0,$$ which is a contradiction. If $r_2 = R_1$ then we get a contradiction in a similar way by
$$0 \ge q'(r_1) = (\lambda_1-\lambda_2) r_1 + \frac{n-1}{r_1} > (\lambda_1-\lambda_2) R_1+ \frac{n-1}{R_1} = 3q'(R_1)
\ge 0.$$
\end{proof}
In the following we will analyze the behavior of $q'$ according to (\ref{EqRic}), considering $r$ and $q$ as two
independent variables. For the sake of a compact notation we will make use of the following abbreviations:
\begin{equation*}
\begin{array}{rclrcl}
p(r) &=& z_1'(r) /z_1(r) \quad \quad & N_y &=& y^2 - n + 1 \cr %
\nu &=& n-2                            & M_y &=& N_y^2/(2y) - \nu^2y/2 \cr %
E &=& \lambda_2-\lambda_1            & Q_y &=& 2y\lambda_1 + EN_yy^{-1} - 2E
\end{array}
\end{equation*}
We further define the function
\begin{equation}\label{EqT}
T(r,y) := -2 p(r) y - \frac{\nu y + N_y}{r} - E r.
\end{equation}
Then we can write (\ref{EqRic}) as
\begin{equation*}
q'(r) = T(r,q(r))
\end{equation*}
The definition of $T(r,y)$ allows us to analyze the Riccati equation for $q'$ considering $r$ and $q(r)$ as independent
variables. For $r$ going to zero, $p$ is $\mathcal{O}(r)$ and thus
$$T(r,y) = \frac{1}{r}\left((\nu+1+y)(1-y)\right) + \mathcal{O}(r) \quad \textmd{for }y\textmd{ fixed}.$$
Consequently,
\begin{equation*}
\begin{array}{rcll}
\lim_{r\rightarrow 0} T(r,y) &=& +\infty  \quad \quad & \textmd{for } 0 \le y < 1 \,\textmd{ fixed,}\cr%
\lim_{r\rightarrow 0} T(r,y) &=& 0 & \textmd{for }\, y = 1 \textmd{ and }\cr%
\lim_{r\rightarrow 0} T(r,y) &=& -\infty & \textmd{for }\, y > 1 \,\textmd{ fixed.}
\end{array}
\end{equation*}
For $r$ approaching $R_1$, the function $p(r)$ goes to minus infinity, while all other terms in (\ref{EqT}) are
bounded. Therefore
$$\lim_{r\rightarrow R_1} T(r,y) = +\infty  \quad \textmd{ for } y > 0 \textmd{ fixed}.$$
The partial derivative of $T(r,y)$ with respect to $r$ is given by
\begin{equation} \label{EqTprime}
T' = \frac{\partial}{\partial r} T(r,y) = -2yp' + \frac{\nu y}{r^2} + \frac{N_y}{r^2}- E.
\end{equation}
In the points $(r,y)$ where $T(r,y) = 0$ we have, by (\ref{EqT}),
\begin{equation}\label{EqPAtTZero}
p|_{T=0} = -\frac {\nu}{2r}-\frac{N_y}{2yr}-\frac{Er}{2y}.
\end{equation}
From (\ref{EquN1}) we get the Riccati equation
\begin{equation}\label{EqRicp}
p'+p^2+\frac{\nu+1}{r}p+\lambda_1-\V=0.
\end{equation}
Putting (\ref{EqPAtTZero}) into (\ref{EqRicp}) and the result into (\ref{EqTprime}) yields
\begin{equation}
T'|_{T=0} = \frac{M_y}{r^2} + \frac{E^2r^2}{2y}+Q_y-2y\V.
\end{equation}
If we define the function
$$Z_y(r) := \frac{M_y}{r^2} + \frac{E^2r^2}{2y}+Q_y-2y\V,$$
it is clear that $T'(r,y) = Z_y(r)$ for any $r,y$ with $T(r,y)=0$. The behavior of $Z_y(r)$ at $r=0$ is determined by
$M_y$. From the definition of $M_y$ we get
\begin{equation}\label{EqMMM}
yM_y  = \frac 12(y^2-1)\cdot[(y-1)-(n-2)]\cdot[(y+1)+(n-2)].
\end{equation}
This implies that
\begin{eqnarray*}
M_y &>& 0 \quad \textmd{for } 0<y<1,\\
M_1 &=& 0.
\end{eqnarray*}
and therefore
\begin{eqnarray*}
\lim_{r\rightarrow 0} Z_y(r) &=& \infty \quad \textmd{for } 0<y<1.
\end{eqnarray*}
\begin{Fact} \label{Fact2}
There is some $r_0>0$ such that $q(r) \le 1$ for $0<r<r_0$ and $q(r_0) < 1$.
\end{Fact}
\begin{proof}
Suppose the contrary, i.e., $q(r)$ first increases away from $r=0$. Then, because $q(0) = 1$ and $q(R) = 0$ and because
$q$ is continuous and differentiable, we can find two points $r_1 < r_2$ such that $\hat q := q(r_1) = q(r_2) > 1$ and
$q'(r_1) > 0 > q'(r_2)$. Even more, we can chose $r_1$ and $r_2$ such that $\hat q$ is arbitrarily close to one.
Writing $\hat q = 1 + \epsilon$ with $\epsilon > 0$, we can calculate from the definition of $Q_y$ that
$$Q_{1+\epsilon} = Q_1 + \epsilon n\left(\lambda_2 - \left(1-2/n\right)\lambda_1\right) + {\mathcal O}(\epsilon^2).$$
The term in brackets can be estimated by
$$\lambda_2 - (1-2/n)\lambda_1 > \lambda_2 - \lambda_1 >0.$$
We can also assume that $Q_1 \ge 0$, because otherwise $q''(0) = \frac{2}{n^2}Q_1 < 0$ and Fact \ref{Fact2} is
immediately true. Thus, choosing $r_1$ and $r_2$ such that $\epsilon$ is sufficiently small, we can make sure that
$Q_{\hat q} > 0.$ We further note, that in view of (\ref{EqMMM}) the constant $M_{\hat q}$ can be positive or negative
(depending on $n$), but not zero because $1<\hat q<2$.

Now consider the function $T(r,\hat q)$. We have $T(r_1,\hat q) > 0 > T(r_2,\hat q)$ and the boundary behavior
$T(0,\hat q) = -\infty$ and $T(R_1,\hat q) = +\infty$. Thus $T(r,\hat q)$ changes its sign at least thrice on
$[0,R_1]$. Consequently, we can find three points $0 < \hat r_1 < \hat r_2 < \hat r_3 < R_1$ such that
\begin{equation}\label{EqZBeh}
Z_{\hat q}(\hat r_1) \ge 0, \quad Z_{\hat q}(\hat r_2) \le 0, \quad Z_{\hat q}(\hat r_3) \ge 0.
\end{equation}
Let us define
$$h(r) = \frac{E^2 r^2}{2 \hat q} - 2 \hat q\V(r).$$
Then
\begin{equation} \label{EqZg}
Z_{\hat q}(r) = \frac{M_{\hat q}}{r^2} + Q_{\hat q} + h(r).
\end{equation}
By condition b) on $\V$, the function $h'(r)$ is concave. Also $h(0) = h'(0) = 0$. We conclude that if $h'(r_0) < 0$ or
$h(r_0) < 0$ for some $r_0> 0$, then $h'(r)$ is negative and decreasing for all $r>r_0$. We will now show that $Z_{\hat
q}$ cannot have the properties (\ref{EqZBeh}), a contradiction that proves Fact \ref{Fact2}:

Case 1: Assume $M_{\hat q} > 0$. Then from $Z_{\hat q}(\hat r_2) \le 0$ we see that
$$-h(\hat r_2) \ge \frac{M_{\hat q}}{\hat r_2^2} + Q_{\hat q} > 0.$$
By what has been said above about $h(r)$, we conclude that $-h(r)$ is a strictly increasing function on $[\hat r_2,\hat
r_3]$. Therefore
$$-h(\hat r_3) > -h(\hat r_2) \ge \frac{M_{\hat q}}{\hat r_2^2} + Q_{\hat q} > \frac{M_{\hat q}}{\hat r_3^2} + Q_{\hat q},$$
such that $Z_{\hat q}(\hat r_3) < 0$, contradicting (\ref{EqZBeh}).

Case 2: Assume $M_{\hat q} < 0$. Then from $Z_{\hat q}(\hat r_1) \ge 0 \ge Z_{\hat q}(\hat r_2)$ follows that $Z_{\hat
q}'(\hat r) \le 0$ for some $\hat r\in [\hat r_1,\hat r_2]$. In view of (\ref{EqZg}) we have $h'(\hat r) < 0$. But this
means by our above concavity argument that $h'(r)$ is decreasing and thus $h'(r) < 0$ for all $r>\hat r$. Then
$Z'_{\hat q}$ is strictly decreasing for $r\ge \hat r$. Together with $ Z_{\hat q}(\hat r_2) \le 0$ and $Z'_{\hat
q}(\hat r) \le 0$ this implies that $ Z_{\hat q}(\hat r_3) < 0$, a contradiction to (\ref{EqZBeh}).
\end{proof} 
\begin{Fact} \label{Fact3}
For all $0\le r\le R_1$ the inequality $q'(r) \le 0$ holds.
\end{Fact} 
\begin{proof}
Assume the contrary. Then there are three points $r_1<r_2<r_3$ in $(0,R_1)$ with $0 < \hat q := q(r_1) = q(r_2) =q(r_3)
< 1$ and $q'(r_1) < 0$, $q'(r_2)>0$, $q'(r_3) < 0$. Consider the function $T(r,\hat q)$, which is equal to $q'(r)$ at
$r_1,r_2,r_3$. Taking into account its boundary behavior at $r = 0$ and $r=R_1$, it is clear that $T(r,\hat q)$ must
have at least the sign changes positive-negative-positive-negative-positive. Thus $T(r,\hat q)$ has at least four zeros
$\hat r_1 < \hat r_2 < \hat r_3 <\hat r_4$ with the properties
$$Z_{\hat q}(\hat r_1) \le 0, \quad Z_{\hat q}(\hat r_2) \ge 0, \quad Z_{\hat q}(\hat r_3) \le 0, \quad Z_{\hat q}(\hat r_4) \ge 0.$$
We also know that $Z_{\hat q}(0) = +\infty$. To satisfy all these requirements, $Z_{\hat q}$ must either have at least
three extremal points where $Z_{\hat q}'$ crosses zero or $Z_{\hat q}$ must vanish on a finite interval. But we have
$$Z'_{\hat q}(r) = -\frac{2M_{\hat q}}{r^3} + \frac{E^2r}{\hat q} -2 \hat q\V'(r),$$
which is a strictly concave function (recall $M_{\hat q} > 0$ for $0<\hat q<1$). A strictly concave function can only
cross zero twice and not be zero on a finite interval, which is a contradiction that proves Fact \ref{Fact3}.
\end{proof} 
Altogether we have shown that $0 < q(r) < 1$ and $q'(r) \le 0$ for all $r\in[0,R]$, proving Lemma
\ref{LemmaMonotonicityOfgAndB}.
\end{proof} 

\section{Proof of Theorem \ref{TheoremPPW}} \label{SectionProof}

\begin{proof}[Proof of Theorem \ref{TheoremPPW}]
We start from the basic gap inequality
\begin{equation} \label{EqGap}
\lambda_2(\Omega,V) - \lambda_1(\Omega,V) \le \frac{\int_\Omega |\nabla P|^2 u_1^2 \di^nr}{\int_\Omega P^2 u_1^2 \di^n
r},
\end{equation}
where $u_1$ is the first Dirichlet eigenfunction of $-\Delta + V$ on $\Omega$ and $P$ is a suitable test function that
satisfies the condition $\int_\Omega Pu_1^2 \di^n r = 0$. We set
\begin{equation}
P_i(\vec r) = g(r) \frac{r_i}{r} \quad \textmd{for } i = 1,2,...,n,
\end{equation}
where
\begin{equation}
g(r) = \left\{\begin{array}{ll} \frac{z_2(r)}{z_1(r)} & \textmd{for } r < R_1\cr %
\lim\limits_{t \uparrow R_1} g(t) & \textmd{for } r \ge R_1.\end{array} \right.
\end{equation}
Here $z_1$ and $z_2$ are the radial parts (both chosen positive) of the first two eigenfunctions of $-\Delta + \V$ on
$S_1$. More precisely, $z_2(r) r_i r^{-1}$ for $i=1,\dots,n$ is a basis of the space of second eigenfunctions. It
follows from the convexity of $r\V$ and the BGM inequality \cite{AB0, BGM} that the second eigenfunctions can be
written in that way.

According to an argument in \cite{AB2} one can always chose the origin of the coordinate system such that $\int_\Omega
P_iu_1^2 \di^n r = 0$ is satisfied for all $i$. Putting the functions $P_i$ into (\ref{EqGap}) and summing over all $i$
yields
\begin{equation} \label{EqDifLambda}
\lambda_2(\Omega,V) - \lambda_1(\Omega,V) \le \frac{\int_\Omega B(r) u_1^2 \di^nr}{\int_\Omega g(r)^2 u_1^2 \di^nr}
\end{equation}
with
\begin{equation*}
B(r) = g'(r)^2 + (n-1)\frac{g(r)^2}{r^2}.
\end{equation*}
By Lemma \ref{LemmaMonotonicityOfgAndB} we know that $B$ is a decreasing and $g$ an increasing function of $r$. Thus,
denoting by $u_1^\star$ the spherically decreasing rearrangement of $u_1$ with respect to the origin, we have
\begin{eqnarray} \label{EqChain1}
\int_\Omega B(r) u_1^2 \di^nr &\le& \int_{\Omega^\star} B^\star(r)\, {u_1^\star}^2 \di^nr\\
&\le& \int_{\Omega^\star} B(r) \,{u_1^\star}^2 \di^nr \le \int_{S_1} B(r)\, z_1^2 \di^nr \nonumber
\end{eqnarray}
and
\begin{eqnarray} \label{EqChain2}
\int_\Omega g(r)^2 u_1^2 \di^nr &\ge& \int_{\Omega^\star} g_\star(r)^2\, {u_1^\star}^2 \di^nr\\
&\ge& \int_{\Omega^\star} g(r)^2 \,{u_1^\star}^2 \di^nr \ge \int_{S_1} g(r)^2\, z_1^2 \di^nr \nonumber
\end{eqnarray}
In each of the above chains of inequalities the first step follows from general properties of rearrangements and the
second from the monotonicity properties of $g$ and $B$. The third step is justified by a comparison result that we
state below and the monotonicity of $g$ and $B$ again. Putting (\ref{EqChain1}) and (\ref{EqChain2}) into
(\ref{EqDifLambda}) we get
$$\lambda_2(\Omega,V) - \lambda_1(\Omega,V) \le \frac{\int_{S_1} B(r)\, z^2 \di^nr}{\int_{S_1} g(r)^2\, z^2 \di^nr} =
\lambda_2(S_1,\V) - \lambda_1(S_1,\V).$$%
Keeping in mind that $\lambda_1(\Omega,V) = \lambda_1(S_1,\V)$, Theorem \ref{TheoremPPW} is proven by this last
inequality.
\end{proof} 

\begin{Lemma}[Chiti Comparison result] \label{LemmaChiti}
Let $u_1^\star$ be the radially decreasing rearrangement of the first eigenfunction of $-\Delta + V$ on $\Omega$ and
$z_1$ the first eigenfunction of $-\Delta + \V$ on $S_1$. Assume both functions to be positive and normalized in
$L^2(\Omega^\star)$. Then there exists an $r_0$ such that
\begin{eqnarray*}
u_1^\star(r) &\le& z_1(r) \quad \textmd{for } r \le r_0 \textmd{ and}\\
u_1^\star(r) &\ge& z_1(r) \quad \textmd{for } r_0 < r \le R_1.
\end{eqnarray*}
\end{Lemma}
\begin{proof} 
By a version of the RFK inequality for Schr\"odinger operators \cite{L} and by domain monotonicity of the first
eigenvalue it is clear that $S_1 \subset \Omega^\star$. This is why we can view $z_1(r)$ as a function in
$L^2(\Omega^\star)$, setting $z_1(r) = 0$ for $r > R_1$.

Both $u_1^\star$ and $z_1$ are positive and spherically symmetric. Moreover, $u_1^\star(r)$ and $z_1(r)$ are decreasing
functions of $r$. For $u_1^\star$ this is clear by definition of the rearrangement. For $z_1$ it follows from a simple
comparison argument using $z_1^\star$ as a test function in the Rayleigh quotient for $\lambda_1$. (Here and in the
sequel we write short-hand $\lambda_1 = \lambda_1(\Omega,V) = \lambda_1(S_1,\V)$.)

We introduce a change of variables via $s = C_n r^n$ and write $u_1^\#(s) \equiv u_1^\star(r)$,  $z_1^\#(s) \equiv
z_1(r)$ and $\V_\#(s) \equiv \V(r)$.
\begin{Fact} \label{FactDE}
For the functions $u_1^\#(s)$ and $z_1^\#(s)$ we have
\begin{eqnarray}
-\frac{\di u_1^\#}{\di s} &\le& n^{-2}C_n^{-2/n}s^{n/2-2} \int_0^s (\lambda_1 - \V_\#(w))\, u_1^\#(w) \di w, \label{EqFact1}\\
-\frac{\di z_1^\#}{\di s} &=& n^{-2}C_n^{-2/n}s^{n/2-2}  \int_0^s (\lambda_1 - \V_\#(w))\, z_1^\#(w) \di w.
\label{EqFact2}
\end{eqnarray}
\end{Fact}
\begin{proof} 
We integrate both sides of $-\Delta u_1 + Vu_1 = \lambda_1 u_1$ over the level set $\Omega_t := \{\vec r\in \Omega:
u_1(\vec r) > t\}$ and use Gauss' Divergence Theorem to obtain
\begin{equation} \label{EqHH}
\int_{\partial\Omega_t} |\nabla u_1| H_{n-1}(\di r) = \int_{\Omega_t} (\lambda_1-V(\vec r))\, u_1(\vec r) \di^n r,
\end{equation}
where $\partial\Omega_t = \{\vec r\in \Omega: u_1(\vec r) = t\}$. Now we define the distribution function $\mu(t) =
|\Omega_t|$. Using the coarea formula, the Cauchy-Schwarz inequality and the classical isoperimetric inequality,
Talenti derives (\cite{T}, p.709, eq. (32))
\begin{equation} \label{EqHH2}
\int_{\partial\Omega_t} |\nabla u_1| H_{n-1}(\di r) \ge - n^2C_n^{2/n} \frac{\mu(t)^{2-2/n}}{\mu'(t)}.
\end{equation}
The left sides of (\ref{EqHH}) and (\ref{EqHH2}) are the same, thus
\begin{eqnarray*}
- n^2C_n^{2/n} \frac{\mu(t)^{2-2/n}}{\mu'(t)} &\le& \int_{\Omega_t} (\lambda_1-V(\vec r))\,u_1(\vec r) \di^n r\\
&\le&  \int_{\Omega_t^\star} (\lambda_1 - V_\star(\vec r))\, u_1^\star(\vec r) \di^n r\\
&\le&  \int_{\Omega_t^\star} (\lambda_1 - \V(\vec r))\, u_1^\star(\vec r) \di^n r\\
&=&  \int_0^{(\mu(t)/C_n)^{1/n}} n C_n r^{n-1} (\lambda_1 - \V(r)) u_1^\star(r) \di r.\\
\end{eqnarray*}
Now we perform the change of variables $r \rightarrow s$ on the right hand side of the above chain of inequalities. We
also chose $t$ to be $u_1^\#(s)$. Using the fact that $u_1^\#$ and $\mu$ are essentially inverse functions to one
another, this means that $\mu(t) = s$ and $\mu'(t)^{-1} = (u_1^\#)'(s)$. The result is (\ref{EqFact1}). Equation
(\ref{EqFact2}) is proven analogously.
\end{proof} 

Fact \ref{FactDE} enables us to prove Lemma \ref{LemmaChiti}. We have $u_1^\#(|S_1|) > z_1^\#(|S_1|) = 0$. Being
equally normalized, $u_1^\star$ and $z_1$ must have at least one intersection on $[0,R]$. Thus $u_1^\#$ and $z_1^\#$
have at least one intersection on $[0,|S_1|]$. Now assume that they intersect at least twice. Then there is an interval
$[s_1,s_2] \subset [0,|S_1|]$ such that $u_1^\#(s) > z^\#(s)$ for $s \in (s_1,s_2)$, $u_1^\#(s_2) = z_1^\#(s_2)$ and
either $u_1^\#(s_1) = z_1^\#(s_1)$ or $s_1=0$. There is also an interval $[s_3,s_4] \subset [s_2,|S_1|]$ with
$u_1^\#(s) < z_1^\#(s)$ for $s \in (s_3,s_4)$, $u_1^\#(s_3) = z_1^\#(s_3)$ and $u_1^\#(s_4) = z_1^\#(s_4)$. Be further
$\tilde s$ the point where $\V_\#(s) - \lambda_1(S_1,\V)$ crosses zero (Set $\tilde s = |S_1|$ if $\V_\#(s) -
\lambda_1$ doesn't cross zero on $[0,|S_1|]$). To keep our notation compact we will write
$$I_a^b[u] =  \int_a^b (\lambda_1-\V_\#(w))\,u(w)\di w.$$
\emph{Case 1:} Assume $\tilde s \ge s_2$. Then $\V_\#(s) - \lambda_1(S_1,\V)$ is negative for $s<s_2$. Set
\begin{equation*}
v(s) = \left\{ \begin{array}{ll} u_1^\#(s) & \textmd{on } [0,s_1] \textmd{ if } I_0^{s_1}[u_1^\#] > I_0^{s_1}[z_1^\#],\cr%
z_1^\#(s) & \textmd{on } [0,s_1] \textmd{ if } I_0^{s_1}[u_1^\#] \le I_0^{s_1}[z_1^\#],\cr %
u_1^\#(s) & \textmd{on } [s_1,s_2]\cr %
z_1^\#(s) & \textmd{on } [s_2,|S_1|]\cr %
\end{array}\right.
\end{equation*}
Using Fact \ref{FactDE}, one can check that then $v(s)$ fulfills the inequality
\begin{equation} \label{Eqv}
-\frac{\di v}{\di s} \le  n^{-2}C_n^{-2/n}s^{n/2-2} \int_0^s (\lambda_1 - \V_\#(s)) v(w) \di w.
\end{equation}
\emph{Case 2:} Assume $\tilde s < s_2$. Then $\V_\#(s) - \lambda_1(S_1,\V)$ is positive for $s\ge s_3$. Set
\begin{equation*}
v(s) = \left\{ \begin{array}{ll} u_1^\#(s) & \textmd{on } [0,s_3] \textmd{ if } I_0^{s_3}[u_1^\#] > I_0^{s_3}[z_1^\#],\cr%
z_1^\#(s) & \textmd{on } [0,s_3] \textmd{ if } I_0^{s_3}[u_1^\#] \le I_0^{s_3}[z_1^\#],\cr %
u_1^\#(s) & \textmd{on } [s_3,s_4]\cr %
z_1^\#(s) & \textmd{on } [s_4,|S_1|]\cr %
\end{array}\right.
\end{equation*}
Again using Fact \ref{FactDE}, one can check that also in this case $v(s)$ fulfills the inequality (\ref{Eqv}).

Now define the test function
$$\Psi(\vec r) = v(C_nr^n) = v(s).$$
Then we use the Rayleigh characterization of $\lambda_1$, equation (\ref{Eqv}) and integration by parts to calculate
\begin{eqnarray*}
\lambda_1 {\int_{S_1} \Psi(\vec r)^2 \di^n r} &<& \int_{S_1} \left(|\nabla \Psi|^2 + \V(\vec r) \Psi^2\right) \di^n r\\
&=& \int_0^{|S_1|} \left( v'(s)^2 n^2 s^{2-2/n} C_n^{2/n} + \V_\#(s) v^2(s)\right) \di s\\
&\le& \int_0^{|S_1|} \left(-v'(s) \int_0^s (\lambda_1-\V_\#(w)) v(w) \di w + \V_\#(s) v^2(s)\right) \di s\\
&=& \int_0^{|S_1|} \left(v(s) (\lambda_1 - \V_\#(s)) v(s) + \V_\#(s) v^2(s) \right) \di s\\
&=& \lambda_1 {\int_{S_1} \Psi(\vec r)^2 \di^n r}.
\end{eqnarray*}
This is a contradiction to our original assumption that $u_1^\#(r)$ and $z_1^\#(r)$ have more than one intersection,
thus proving Lemma \ref{LemmaChiti}.
\end{proof} 

\section{Proof of Theorem \ref{TheoremMonotonicity}} \label{SectionProof2}

\begin{proof}[Proof of Theorem \ref{TheoremMonotonicity}]
The first eigenfunction of $-\Delta + V$ on $B_R$ is radially symmetric and will be called $z_1(r)$. Further, a
standard separation of variables and the Baumgartner-Grosse-Martin \cite{BGM, AB3} inequality imply that we can write a
basis of the space of second eigenfunctions in the form $z_2(r) \cdot r_i \cdot r^{-1}$. The radial parts $z_1$ and
$z_2$ of the first and the second eigenfunction, which we assume to be positive, solve the differential equations
\begin{eqnarray}
-z''_1(r) - \frac{n-1}{r} z'_1(r) + \left(V(r) -\lambda_1\right) z_1(r) &=& 0, \label{EqDEs}\\
-z''_2(r) - \frac{n-1}{r} z'_2(r) + \left(\frac{n-1}{r^2} + V(r)- \lambda_2\right) z_2(r) &=& 0 \nonumber
\end{eqnarray}
with the boundary conditions
\begin{equation}\label{EquM3}
z'_1(0) = 0, \quad z_1(R) = 0, \quad z_2(0) = 0, \quad z_2(R) = 0.
\end{equation}
We define the rescaled functions $\tilde z_{1/2}(r) = z_{1/2}(\beta r)$. Putting $\beta r$ (with $\beta>0$) instead of
$r$ into the equations (\ref{EqDEs}) and multiplying by $\beta^2$ yields the rescaled equations
\begin{eqnarray*}
-\tilde z''_1(r) - \frac{n-1}{r} \tilde z'_1(r) + \left(\beta^2 V(\beta r) -\beta^2 \lambda_1\right) \tilde z_1(r) &=& 0, \nonumber\\
-\tilde z''_2(r) - \frac{n-1}{r} \tilde z'_2(r) + \left(\frac{n-1}{r^2} + \beta^2 V(\beta r)- \beta^2 \lambda_2\right)
\tilde z_2(r) &=& 0. \nonumber
\end{eqnarray*}
We conclude that $\tilde z_1$ and $\tilde z_2$ are the radial parts of the first two eigenfunctions of $-\Delta +
\beta^2 V(\beta r)$ on $B_{R/\beta}$ to the eigenvalues $\beta^2 \lambda_1$ and $\beta^2\lambda_2$. Consequently, if we
replace $R$ by $R/\beta$ and $V(r)$ by $\beta^2 V(\beta r)$, then the ratio $\lambda_2/\lambda_1$ doesn't change.

For the rest of this section we shall write $\lambda_{1/2}(R,V)$ instead of $\lambda_{1/2}(B_R,V)$. We also fix two
radii $0 < R_1 < R_2$ and let $\rho(\beta)$ for $\beta > 1$ be the function defined implicitly by
\begin{equation} \label{EqLam}
\lambda_1(\rho(\beta), V(r)) = \lambda_1(R_2/\beta, \beta^2 V(\beta r)).
\end{equation}
Then we have $\rho(1) = R_2$. By domain monotonicity of $\lambda_1$ and because $V(r)$ is increasing and positive we
see that the right hand side of (\ref{EqLam}) is increasing in $\beta$. Therefore, again by domain monotonicity,
$\rho(\beta)$ must be decreasing in $\beta$. One can also check that $\rho(\beta)$ is a continuous function and that
$\rho(\beta)$ goes to zero for $\beta \rightarrow \infty$. Thus we can find $\beta_0 > 1$ such that $\rho(\beta_0) =
R_1$. Then we can apply Theorem \ref{TheoremPPW}, with $B_{R_2/\beta_0}$ for $\Omega$ and  $B_{\rho(\beta_0)}$ for
$S_1$, as well as $\beta_0^2 V(\beta_0 r)$ for $V$ and $V(r)$ for $\V$, to get
\begin{equation} \label{EqLam2}
\lambda_2(R_2/\beta_0, \beta_0^2 V(\beta_0 r) \le \lambda_2(\rho(\beta_0), V(r)) = \lambda_2(R_1, V(r)) .
\end{equation}
But by what has been said above about the scaling properties of the problem, we have
\begin{equation} \label{EqLam3}
\frac{\lambda_2(R_2/\beta_0, \beta_0^2 V(\beta_0 r))}{\lambda_1(R_2/\beta_0, \beta_0^2 V(\beta_0 r))} =
\frac{\lambda_2(R_2, V(r))}{\lambda_1(R_2, V(r))}.
\end{equation}
Combining (\ref{EqLam}) for $\beta=\beta_0$, (\ref{EqLam2}) and(\ref{EqLam3}), we get
\begin{equation} \label{EqLam4}
\frac{\lambda_2(R_1,V(r))}{\lambda_1(R_1, V(r))} \ge \frac{\lambda_2(R_2, V(r))}{\lambda_1(R_2, V(r))}.
\end{equation}
Because $R_1$ and $R_2$ were chosen arbitrarily, this proves Theorem \ref{TheoremMonotonicity}.
\end{proof}

\section{Proof of Theorem \ref{TheoremMonotonicity2}} \label{SectionProof3}

Before we prove Theorem \ref{TheoremMonotonicity2} we need to state the following technical Lemma:
\begin{Lemma}\label{LemmaAlgebraic}
Be $a,b,c,d > 0$ with $a\ge b$, $d\ge b$ and $\frac a b < \frac cd$. Then
\begin{equation}
\frac{a+x}{b+x} < \frac{c+x}{d+x}
\end{equation}
holds for any $x>0$.
\end{Lemma}
\begin{proof}
Define the function
$$f(x) := \frac{c+x}{d+x} - \frac{a+x}{b+x}.$$
then $f(0) > 0$. A straightforward calculation shows that $f$ has exactly one zero at
$$x_0 = -\frac{bc-ad}{b+c-a-d}.$$
The numerator $bc-ad$ in the expression for $x_0$ is positive because of the condition $\frac a b < \frac cd$. For the
denominator we get
$$b+c-a-d > c+b-\frac{bc}{d}-d = \frac{(d-b)(c-d)}{d} \ge 0.$$
This means that $x_0<0$, such that $f(x)>0$ for all $x>0$.
\end{proof}

\begin{proof}[Proof of Theorem \ref{TheoremMonotonicity2}]
Choose some $x > 0$. From Theorem \ref{TheoremMonotonicity} we know that
$$\frac{\lambda_2(B_{R+x},r^2)}{\lambda_1(B_{R+x},r^2)} < \frac{\lambda_2(B_{R},r^2)}{\lambda_1(B_{R},r^2)} \quad \textmd{for }x>0.$$ %
Moreover, $\lambda_1(B_R,r^2) \ge \lambda_1(B_{R+x},r^2)$ and $\lambda_2(B_{R+x},r^2) > \lambda_1(B_{R+x},r^2)$. Thus
we can apply first (\ref{EqCon}), then Lemma \ref{LemmaAlgebraic} and then (\ref{EqCon}) again, to get
\begin{equation*}
\frac{\lambda^+_2(B_{R+x})}{\lambda^+_1(B_{R+x})}  = \frac{\lambda_2(B_{R+x},r^2) + n}{\lambda_1(B_{R+x},r^2) + n}
< \frac{\lambda_2(B_{R},r^2) + n}{\lambda_1(B_{R},r^2) + n} = \frac{\lambda^+_2(B_{R})}{\lambda^+_1(B_{R})}.%
\end{equation*}
\end{proof}


\end{document}